# Realization and Properties of Biochemical-Computing Biocatalytic XOR Gate Based on Signal Change

Vladimir Privman, Jian Zhou, Jan Halámek and Evgeny Katz*

*Department of Chemistry and Biomolecular Science, and*
*Department of Physics, Clarkson University, Potsdam, NY 13676*

**\***Corresponding author:
E-mail: ekatz@clarkson.edu; Tel.: +1 (315) 268-4421; Fax: +1 (315) 268-6610

## ABSTRACT

We consider a realization of the **XOR** logic gate in a system involving two competing biocatalytic reactions, for which the logic-**1** output is defined by these two processes causing a change in the optically detected signal. A model is developed for describing such systems in an approach suitable for evaluation of the analog noise amplification properties of the gate and optimization of its functioning. The initial data are fitted for gate quality evaluation within the developed model, and then modifications are proposed and experimentally realized for improving the gate functioning.

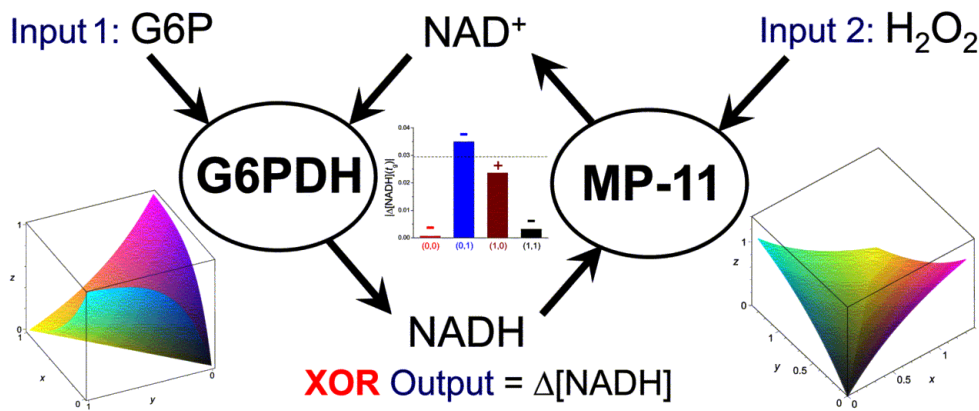

1.      Introduction

Rapid development of chemical computing,[1-7] i.e., information processing encoded in chemical reaction kinetics, in turn a subarea of unconventional computing,[8,9] resulted in realizations of various chemical systems carrying out Boolean logic gates, such as **AND**,[10,11] **OR**,[12] **NAND**,[13,14] **NOR**,[15-18] **INHIB**,[19-22] **XOR**,[23-26] etc. Recently emerging field of biochemical computing[27-29] has resulted in systems for Boolean logic gate functions, as well as few-gate networks, utilizing biomolecular processes, such as those based on proteins/enzymes,[30-38] antigens/antibodies,[39,40] DNAzymes,[41,42] DNA,[43,44] RNA[45-49] and even whole biological cells.[50] One of the motivations for the interest in biocomputing systems has been their potential applications for novel multi-signal responsive biosensors[51-53] and bioactuators[54,55] logically processing complex patterns of biochemical signals, with promise of biomedical applications.[56-59]

Among all Boolean logic operations, **XOR** (e**X**cluded **OR**) gate is the most difficult for the chemical realization. This gate should produce **0** output when the inputs are applied at zero levels (input combination **0**,**0**). Each input signal applied separately should result in the system activation (giving output **1** upon input combinations **0**,**1** and **1**,**0**). However, simultaneously applying both inputs should keep the system inactive (resulting in the output **0** when the inputs **1**,**1** are applied). Note that **AND** logic gates, for instance, respond to two chemical stimuli in a natural for chemistry way generating the product only in the presence of both chemicals in the system,[10,11] while **OR** gates can be represented by two parallel chemical pathways resulting in similar products in the presence of any or both chemical stimuli.[12] **NAND** / **NOR** logic gates represent inversion of the **AND** / **OR** functions whereby a certain chemical disappears when the gates are active.[13-18] **INHIB** logic gate is also easy for chemistry to mimic since it represents inhibition of the product formation in the presence of a certain chemical input, which is also rather natural for chemical systems.[19-22] However, the **XOR** function should result in a chemical output only in the presence of any one of two inputs applied separately, but the product formation should be inhibited in the case of simultaneous presence of both inputs.[23-26] This is quite unusual for chemical kinetics because two potentially reactive species should cancel each other when they both appear in the system.

In non-biochemical systems, the **XOR** function has usually been realized by mutual negation of the chemical input signals, e.g., by neutralization of acid and base being used as two inputs.[23,24] It should be noted that this rather simple way of operation is based on the neutralization of the inputs prior to their reaction with the logic gate system. In more sophisticated supra-molecular systems input signals resulted in the conversion of a chemical complex to another structure (reflected for example by changes in optical absorbance or fluorescence) upon binding each chemical input separately at different sites of the complex, while binding both of them together resulted in no changes in the optical properties, thus mimicking **XOR** logic gate.[25,26] Biomolecular realization of **XOR** gate can be based on specific biorecognition and biocatalytic properties of biochemical systems, using DNA[48] and enzymes,[60-62] respectively. In DNA-based **XOR** gates single-strand oligonucleotides are used as system activating input signals, while their hybridized double-strand derivatives do not activate the logic gate.

In enzyme-based logic gates oppositely directed biocatalytic reactions can be activated by input signals. Separate application of each chemical input activates one of the reactions resulting in the unbalancing of the system and producing the output signal, while simultaneous addition of the both inputs would activate both competing reactions, thus keeping a balance in the system and producing no chemical output.[38,60-62] It should be noted that this kind of **XOR** gate produces oppositely directed chemical changes in the systems (for example increase or decrease in the concentration of a certain chemical) upon activation with different input signals. In order to fulfill the **XOR** gate definition, the output signals in these gates were defined as absolute values of the concentration changes (or absolute values of the concentration dependent parameters, such as optical[60-62] or electrochemical output signals[38]). The latter type of a system is considered in this work.

Despite the difficulties in the chemical realization of the **XOR** logic gate, specific interest in it is based on its importance for molecular computing systems where **XOR** gate is a part of basic computing elements: half-adder/half-subtractor[63-66] or full-adder/full-subtractor.[67,68] These basic arithmetic functions were realized using DNA-based[69] and enzyme-based[70] **XOR** gates.

**XOR** is also an interesting function for other reasons. It is a part of certain reversible gates (**CNOT**, **CCNOT**) that have been extensively studied in other unconventional computing realizations.[71-76] Furthermore, its truth table, given in Scheme 1, suggests that a response-surface function of a system that yields **XOR** at the logic-point inputs **00**, **01**, **10**, **11**, may have a *saddle shape* passing through the two output **0**s at **00** and **11**, but also through the output **1**s at **01** and **10**. This makes the study of the noise scaling as a signal is processed by this gate interesting, and also suggests more than a single pattern of behavior (different shapes of the "saddle"). Thus, various types of **XOR** gates may have to be considered separately as noise handling components of biochemical logic "networks." One such study is carried out in the present work. Generally, however, we expect that the **XOR** gate is more noisy that **AND**, **OR** and other gates with less "structured" (means, less sloped except at selected points) response surfaces connecting their logic points. Therefore, while the study of the noise handling properties upon transmission through an **XOR** gate is interesting, we expect up front that this gate will be a candidate for inclusion in networks only provided noise suppressing elements (filters) are utilized, rather than directly optimizable for low-noise functioning.

The article is organized as follows. Our experimental system, outlined in Scheme 1, is presented in Section 2. Section 3 reports a model suitable for the specific type of **XOR** functioning considered here. Finally, Section 4 presents results, their model analysis, optimization, as well as discussion and concluding remarks.

**2. Experimental: XOR Gate Based on Signal Change**

*Chemicals and Materials.* Glucose-6-phosphate dehydrogenase (G6PDH) from *Leuconostoc mesenteroides* (E.C. 1.1.1.49), microperoxidase-11 (MP-11) sodium salt ~90% (HPLC), D-glucose-6-phosphate sodium salt (G6P), β-nicotinamide adenine dinucleotide sodium salt (NAD$^+$) (≥95%), and β-nicotinamide adenine dinucleotide reduced dipotassium salt (NADH) were purchased from Sigma-Aldrich and used without further purification. H$_2$O$_2$ (30% w/w) was purchased from Fisher. Ultrapure water (18.2 MΩ·cm) from NANOpure Diamond (Barnstead) source was used in all of the experiments.

*Composition and mapping of the **XOR** System.* The "machinery" of the **XOR** gate was composed of the enzyme: G6PDH (0.1 unit·mL$^{-1}$) and biocatalyst: MP-11 (8.6 µM) operating together with NADH (100 µM) and NAD$^+$ (100 µM) in 50 mM phosphate buffer, pH 7.4. Another set of mapping experiments was performed with lower enzyme and biocatalyst concentrations: G6PDH (0.05 unit·mL$^{-1}$), MP-11 (0.86 µM), following optimization suggested by modeling reported in Sections 3-4. The enzyme substrate G6P and the biocatalyst oxidizer H$_2$O$_2$ were used as the input signals. Logic value **0** for the input signals was defined as the absence of G6P and H$_2$O$_2$ in the reacting solution, while their presence at the concentrations of 1 mM and 50 µM, respectively, was defined as input values **1**. In the gate-response mapping experiments the input signals were applied at variable concentrations, G6P: 0, 0.25, 0.50, 0.75, 1.00 and 1.25 mM, and H$_2$O$_2$: 0, 12.5, 25.0, 37.5, 50.0 and 62.5 µM. The response matrix of 36 experimental points was obtained for these input concentrations.

*Optical Measurements.* Absorbance, A, measurements were performed using a UV-2401PC/2501PC UV-visible spectrophotometer (Shimadzu, Tokyo, Japan) at (37 ± 0.2) °C. The reactions took place in a 1 mL poly(methyl methacrylate), PMMA, cuvette. The change of the NADH absorbance was monitored at $\lambda = 340$ nm.[77] A selection of our absorbance data is shown in Figure 1. The sampling time, $t_g$, was selected to obtain a well-defined **XOR** function, as detailed in Section 3. The absorbance values were converted to the concentration of NADH using its molar extinction coefficient, $\varepsilon_{340} = 6.22 \cdot 10^3$ M$^{-1}$·cm$^{-1}$.[77] The **XOR** output signal was defined as the absolute value of the change in the NADH concentration during the gate function. Since some other chemicals, most notably MP-11 and NAD$^+$, can contribute, to a limited extent, to the absorbance at $\lambda = 340$ nm, we used the change in the absorbance as compared to that at time 0 for input (**0,0**) to estimate the difference $\Delta$[NADH]($t_g$) = [NADH]($t_g$) – [NADH](0), rather than directly subtracting the actual [NADH](0) value.

*Parameter Selection for the **XOR** Gate.* Generally, the system of biocatalytic reactions shown in Scheme 1, will not yield the **XOR** binary function unless the experimental parameters are properly selected. Here we consider the initial-time, $t = 0$, concentration [G6P](0) as Input 1, with the reference binary logic values 0 and [G6P]$_{max}$. Similarly, [H$_2$O$_2$](0) is Input 2, with the logic values **0** and [H$_2$O$_2$]$_{max}$. The **XOR** output, |Δ[NADH]($t_g$)|, is calculated from the change in the absorbance, as the absolute value of the difference Δ[NADH]($t_g$) at a certain "gate time," $t_g > 0$, with respect to the initial value at $t = 0$ for input (**0,0**), as mentioned above. The inputs are in principle determined by the utilization of the gate in specific applications, whereas the logic-**0** and **1** values of the output, here assumed 0 and |Δ[NADH]|$_{max}$, are largely set by the gate-function itself (up to uncertainty due to intrinsic noise). However, the relative "activities" of the two branches of the reaction must be adjusted to yield the **XOR** function, depending on the initial ratio of [NADH](0)/[NAD$^+$](0). This point will be explained by the model of the next section, as will be our choice of this ratio at value **1**.

Furthermore, the quality of the XOR function depends on the choice of the gate time, $t_g$. The choice of $t_g$ will also be to a certain extent explained by the model in Section 3, which is, however, a simplified phenomenological description of the biocatalytic kinetics. The gate times marked by the vertical lines in Figure 1, were selected to have both logic-**0** output values approximately equal to 0, and for both logic-**1** outputs — approximately equal to each other absolute values of the deviation from 0. However, we also sought gate times for which the outputs for input pairs multiplied by equal factors (additional data also shown in the figure), are close to 0. This will be explained in the next section. The choice of the gate times is not unique, and the quality of the approximation of the **XOR** by the output signals is not ideal, see Figure 2, but it definitely compares favorably to a typical such built-in "intrinsic-noise" spread/shift of logic-point values for gates reported in the chemical and biochemical computing literature. The actual values for our choice of $t_g = 150$ sec and 215 sec for the two experiments shown in the top and bottom panels of Figure 1, respectively, are shown as bar charts in Figure 2.

## 3. Model of the XOR Response Surface

As described in the preceding section, in order to explore the noise scaling properties[78-81] of the realized **XOR** gate we map out the output of the biocatalytic processes at input values other than the logic **0**s and **1**s. Specifically, we define the rescaled logic-range variables

$$x = [\text{G6P}](0) / [\text{G6P}]_{\max},$$

$$y = [\text{H}_2\text{O}_2](0) / [\text{H}_2\text{O}_2]_{\max}, \quad (1)$$

$$z = |\Delta[\text{NADH}](t_g)| / |\Delta[\text{NADH}]|_{\max},$$

in terms of which the gate-response function $z(x,y)$ can be considered, with $0 \leq x, y, z \leq 1$. Our experiments provide a mapping of this function for 36 combinations $x, y = 0, ..., 1.25$, within and somewhat beyond the "logic" range. The behavior of the function $z(x,y)$ near the logic points $(x,y) = (0,0), (0,1), (1,0), (1,1)$ determines[78] the degree to which the gate amplifies or suppresses analog noise in the inputs. There are other sources of noise in biomolecular gates and in their networks with other gates and non-binary elements. For example, the deviations of the output from the selected logic-**0** or **1** binary values, see Figure 2, are part of the "intrinsic" noise in the values of the function $z(x,y)$ itself, which ideally should be exactly at $z = 0, 1, 1, 0$ for $xy = 00$, 01, 10, 11 input combinations, respectively.

Most biochemical gates studied for their analog-noise amplification properties, have a smooth, convex function $z(x,y)$. The noise-spread scaling factor upon noisy signals passing through the gate, can then be estimated simply as the slope, $|\vec{\nabla} z(x,y)|$ at each of the four logic points. The largest of the slopes has typically been found larger than 1 for convex **AND** gate response surfaces studied.[78] Therefore, such gates typically amplify noise, by a significant factor, e.g., 300–500%. However, the gate-response function $z(x, y;...)$, in addition to the scaled inputs $x$ and $y$, also depends parametrically on other variables, here shown as …, such as the initial concentrations of the biocatalysts and other chemicals that can be adjusted. For **AND** gates the experience has been[27,78] that noise amplification can be reduced to approximately 120% (factor ~ 1.2). However, this direct gate optimization is difficult because large changes in the controllable concentrations and other parameters are needed. Other approaches have been tried, such as non-smooth surfaces and partially sigmoidal ones. These have been limited to specific examples,[27,80,81] and the present understanding seems to be that, only network-level optimization (filtering, etc.) with the help of additional network elements, can fully control analog noise proliferation for useful applications of complex biochemical logic circuitry. Ultimately, for larger networks, of more than about 10 gates, digital error correction by redundancy will also be required.[78,81]

Here we aim at a study of the noise scaling properties of the realized **XOR** gate. In order to develop a model for the function $z(x,y;...)$, we could attempt a detailed kinetic description of the processes biocatalyzed by G6PDH and MP-11, see Scheme 1. The mechanisms of action of both biocatalysts, especially G6PDH, is, however, rather complicated,[82-84] involving several complex pathways and a possible allostericity.[80,85] Detailed modeling would require introduction of several rate constants, most of which are not known from the literature, and is not practical based only on the available data for the present **XOR** gate realization. The other extreme has been to rely on purely phenomenological fitting forms[79,86] for $z(x,y;...)$. The latter approach, tested for **AND** gates, including a small network,[86] introduces a couple of ad-hoc parameters — those denoted by … in our notation for $z(x,y;...)$ — which can be fitted from the data, but their relation to the adjustable physical or chemical conditions of the experiment is not straightforward.

Here we adapt an intermediate approach. Due to complexity of the actual kinetics we do not attempt detailed modeling. However, we do retain a more direct relation of the rate parameters introduced to the actual processes involved, which is useful for gate-functioning optimization, as will be illustrated in this and the next section. Specifically, we consider the two branches of the **XOR** scheme, converting the two "competing" chemicals: NADH, of concentration of be denoted $P(t)=[\text{NADH}](t)$, and NAD$^+$, of concentration $S(t)=[\text{NAD}^+](t)$, into one another:

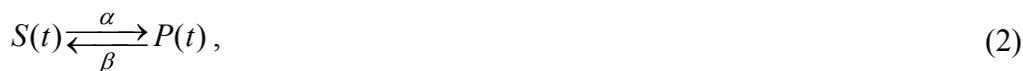

$$S(t) \underset{\beta}{\overset{\alpha}{\rightleftarrows}} P(t), \qquad (2)$$

where the rate constants, $\alpha, \beta$, will be the two phenomenological quantities that represent the overall activity of the respectively G6PDH and MP-11 branches of the processes depicted in Scheme 1. Specifically, we expect that $\alpha$ is roughly proportional to the initial concentrations of Input 1, [G6P](0), and the enzyme, [G6PDH](0), whereas $\beta$ is similarly related to Input 2, [H$_2$O$_2$](0), and the biocatalyst, [MP-11](0). This is obviously a rather crude model, because it lumps complicated kinetics into only a single parameter for each branch. Furthermore, even if the model is taken at face value, conclusions such as the approximate proportionalities just mentioned, can be questioned. We already mentioned a possible allostericity[80,85] of G6PDH, implying nonlinear dependence on its substrate concentration(s). For the other branch, hydrogen peroxide can in fact inhibit MP-11, though we are working well below the relevant concentrations.[87,88]

Despite its obvious gross limitations as a realistic model of the biocatalytic processes involved, the present approach has advantages that make it practical for our purposes. It limits the number of fit parameters to reasonable (just two) for the quality of the available data. In addition, we note that we are only interested in a rather approximate description of the whole response surface, which will capture its general shape features and specifically the slope values near the four logic points and their relative variation as functions of the adjustable quantities. Finally, the present model provides a direct prescription for varying the introduced parameters $\alpha, \beta$, by adjusting the main controlled quantities of the gate function: the initial biocatalyst concentrations. We expect that variation of the latter will roughly translate into proportional variation of the respective rate constants.

The rate equations corresponding to Eq. (2) are easy to formulate and solve:

$$\frac{dP(t)}{dt} = -\frac{dS(t)}{dt} = \alpha S(t) - \beta P(t),$$

$$P(t_g) - P(0) = \frac{\alpha S(0) - \beta P(0)}{\alpha + \beta}[1 - e^{-(\alpha+\beta)t_g}],$$

(3)

where the result is here written in the form the absolute value of which is our **XOR** signal at $t_g$. We note that, since the parameters $\alpha, \beta$ are assumed proportional to the inputs, then their values at the logic-**1** inputs ([G6P]$_{max}$, [H$_2$O$_2$]$_{max}$) can be identified: $\alpha_{max}, \beta_{max}$. Eq. (1) then suggest that $x = \alpha / \alpha_{max}$ and $y = \beta / \beta_{max}$. The calculation of $z$ in this notation will be addressed later. It is important to emphasize that while Eq. (3) gives signal 0 at inputs (0,0), it should also give output 0 at inputs $(\alpha_{max}, \beta_{max})$. Otherwise, we will not have the **XOR** function. Furthermore,

– 9 –

the positive output signal at inputs $(\alpha_{\max}, 0)$ and the negative signal obtained at $(0, \beta_{\max})$, should have magnitudes equal to one another, because otherwise the gate will not mimic the binary **XOR** after taking the absolute value for the final signal definition. This suggests that the parameters of the model are not arbitrary but must satisfy the following relations:

$$\frac{S(0)}{P(0)} = \frac{\beta_{\max}}{\alpha_{\max}} = \frac{1-e^{-\beta_{\max}t_g}}{1-e^{-\alpha_{\max}t_g}}. \tag{4}$$

If our model is taken literally as the precise description of the biocatalytic processes involved in the gate dynamics, then Eq. (4) imply that for an arbitrary choice of the gate time $t_g$, we have to adjust the activities (balance the amounts of the biocatalysts) in the two branches, and also adjust the initial quantities of the two chemicals that are being interconverted into one another, to have

$$\alpha_{\max} = \beta_{\max} \quad \text{and} \quad S(0) = P(0). \tag{5}$$

In reality, of course, the present model is approximate and therefore, these requirements are not fully accurate. Indeed, we selected equal initial values of the concentrations $P(0)$ = [NADH](0) and $S(0)$ = [NAD$^+$](0) to have a well-approximated **XOR** function. However, the relation between the variables $\alpha_{\max}$, $\beta_{\max}$ and $t_g$ is in reality likely more complicated than predicted in Eq. (5). Indeed, the (**1,1**) data in particular, in Figure 1, shows the time dependence which is obviously not monotonic, as predicted by relations of the type of Eq. (3). Therefore, the model predictions, and the degree to which the **XOR** function can be reproduced by the present system, are both limited. In fact, our specific values of $t_g$ selected, marked in Figure 1, are those for which a compromise is reached in the degree to which the system approximates all the aforementioned **XOR** requirements. Furthermore, the form of Eq. (3) suggests that the signal should actually vanish for this type of **XOR** also when the inputs are equal fractions of the maximum values: $\alpha/\beta = \alpha_{\max}/\beta_{\max}$. Indeed, as seen in Figure 1, the data sets for such fractional inputs approximately all intersect close to zero signal in the vicinity of the selected $t_g$. Thus, the reference gate times selected, were those that approximately yield output **0** for several equal inputs, varying from both inputs being zero to both inputs assuming values 1 and beyond.

Furthermore, the inputs **01** and **10** were expected to yield approximately equal in absolute value signals based on the absorbance change, at $t_g$.

With these reservations in mind both with regards to the precision of the model and to the accuracy with which the selected system actually yields **XOR**, we can utilize the present model for a semi-quantitative data fit to attempt to optimize the initial, randomly selected (for experimental convenience) parameter values (specifically, the biocatalyst concentrations which control the "activity" of the two branches). This optimization, carried out in Section 4, yielded the second set of experiments reported in Figures 1 and 2.

Within the framework of the model, if all the conditions are exactly satisfied, we can now divide the absolute value of the difference signal in Eq. (3), by the maximal signal, and introduce the rescaled variables, to get the final expression for the gate-response function,

$$z(x,y) = \frac{|x-y|}{x+y} \frac{1-e^{-a(x+y)}}{1-e^{-a}}, \quad \text{where} \quad a \equiv \alpha_{max} t_g \, (= \beta_{max} t_g). \tag{6}$$

Thus, within the present model the **XOR** gate response surface depends on a single dimensionless parameter, $a \geq 0$. Note that as $a \to 0$, we have $z(x,y) \to |x-y|$. Figure 3 shows the function in Eq. (6) for illustrative values $a = 4$ and ¼. The function is smooth except for the crease at $z = 0$, along the diagonal $x = y$. The latter is due to taking the absolute value, and we know from other studies[79] (for an **AND** gate) that such a line of non-smooth behavior does not invalidate the property that, for narrow enough noise distribution in the inputs, the slope of the function (without the |…|) in Eq. (6) gives a good measure of the noise amplification factor as the signal is converted into the output. These slopes are plotted in Figure 4.

Figure 4 indicates that the largest slope is always greater than approximately 1.4, which means that **XOR** gates of this type always amplify analog noise by adding at least 40% to the relative width of the signal distribution. We note that a similar estimate for **AND** gates was close to 20%,[78] which confirms our expectation that **XOR**s — gates with two diagonally positioned outputs **0** and two outputs **1** — are generally noisier than gates with only a single output value different from the other three outputs. Figures 3 and 4 indicate that the optimal shape for reduced

– 11 –

analog noise amplification for this type of **XOR** gates — those with the "saddle pass" depressed low towards the *xy* plane (in our case, the extreme: touching the *xy* plane as the diagonal crease) — is made of two flat triangles, obtained in the limit of small *a*. This, however, brings out another problem: such linear response surfaces are usually obtained when the concentrations of most of the reactants are generally made small (to decrease the activities of all the reaction pathways) or the gate time is very small. Then the separation between the physical values for logic-**0** and **1** outputs will also be reduced, resulting in another source of noise: The intrinsic noise due to inaccuracies in the gate realization and fluctuations in the response function itself might become large on the relative scale (i.e., it will be difficult to distinguish between the logic-**0** and **1** signals). Thus, **XOR** gate optimization is generally more challenging than that for **AND** gates.

### 4. Analysis of the Experimental Data, and Discussion

Our first set of the reaction parameters was selected based primarily on the experimental convenience. We did satisfy one of the model requirements: $[NAD^+](0) = [NADH](0)$, which are expected to lead to better conformity of the resulting output values with the **XOR** function. However, the second requirement, that of making the "activities" of the two biocatalytic branches approximately equal, $\alpha_{max} = \beta_{max}$, is less straightforward not only because the proportionality constants relating these quantities to concentration are not known, but also because the applicability of our phenomenological model itself is at best semi-quantitative. We will return to this condition shortly.

In order to cast our data, e.g., Figure 1, at $t_g$ in terms of the "logic" value ranges, we kept the choice of the physical 0 as the logic-**0**, despite the fact that the actual gate realizations give small positive outputs (Figure 2). For logic-**1**, we took the arithmetic means of the two positive values obtained at the appropriate logic points, **01** and **10** (Figure 2). Our original data set at $t_g$, is displayed in Figure 5. The data were fitted by the form suggested by Eq. (6). The estimate for the adjustable parameter was $a = 1.06$. This suggests that the original gate is rather noisy: at this value of $a$, the noise amplification factor is 229% (this is outside the range of Figure 4), i.e., the noise level is more than doubled due to the gate function. Furthermore, the quality of the root-mean-squares fit of the data is not that great. This is an indirect indication that the model is not describing the data well. Indeed, consideration of the time-dependence of the (**0**,**1**) vs. (**1**,**0**) curves in the top panel of Figure 1, suggests that the MP-11 branch of the reaction is much more "active" (faster onset of saturation) than the G6PDH branch, i.e., that the original parameter selection was quite far from satisfying the requirement $\alpha_{max} = \beta_{max}$ for a quality **XOR** functioning.

The above considerations suggest that an improved set of parameters should be selected whereby the activities of both branches of the process are lowered, but the one for MP-11 should be reduced by a significantly larger factor. Indeed, our modified experiment, for which the "logic" data and its model fit are shown in Figure 6, had the initial concentration of G6PDH reduced by a factor of 2, whereas that of MP-11 reduced by a factor of 10. While the resulting **XOR** function is still not fully symmetric (Figure 2), the time scales of the variation of the (**0**,**1**) vs. (**1**,**0**) curves in the bottom panel of Figure 1, are more comparable, which suggests that the condition $\alpha_{max} = \beta_{max}$ is closer satisfied. Furthermore, the fitted value, $a = 0.459$, now corresponds to the noise amplification factor of 176%, which is a significant improvement. Note that the optimal value for this type of **XOR** gate is approximately 140%, but to achieve it, the activity of the reaction branches would have to be reduced too much, resulting in very small signals. Indeed, our optimization already reduced the output signal strength by approximately a factor of 2, as can be seen in Figure 2.

In summary, we developed an approach to analyze and optimize the noise-scaling properties of **XOR** gates of the "signal change" type.[38,60-62] Our general conclusions are that, such gates are actually somewhat noisier and more difficult to optimize than earlier studied **AND** gates. We emphasize that, as described in the Introduction, other types of **XOR** gate are possible, and their noise-handling properties should be studied in the future. For the presently considered gates, the smallest possible noise amplification factors of ~ 1.4 are difficult to achieve, because the regime required is that of weak output signals. However, we demonstrated experimentally that factors ~ 1.8 are feasible with reasonable degree of modeling-enabled optimization of the gate-parameter choices.

**Acknowledgements**

The authors thank Dr. V. Bocharova for helpful scientific input and collaboration, and M. A. Arugula for preliminary experiments and technical assistance. The authors gratefully acknowledge research funding by the NSF (grant CCF-0726698).

# References

1. De Silva, A. P.; Uchiyama, S.; Vance, T. P.; Wannalerse, B. *Coord. Chem. Rev.* **2007**, *251*, 1623-1632.
2. De Silva, A. P.; Uchiyama, S. *Nature Nanotechnol.* **2007**, *2*, 399-410.
3. Szacilowski, K. *Chem. Rev.* **2008**, *108*, 3481-3548.
4. Credi, A. *Angew. Chem. Int. Ed.* **2007**, *46*, 5472-5475.
5. Pischel, U. *Angew. Chem. Int. Ed.* **2007**, *46*, 4026-4040.
6. Pischel, U. *Austral. J. Chem.* **2010**, *63*, 148-164.
7. Andreasson, J.; Pischel, U. *Chem. Soc. Rev.* **2010**, *39*, 174-188.
8. *Unconventional Computation. Lecture Notes in Computer Science.* Calude, C. S.; Costa, J. F.; Dershowitz, N.; Freire, E.; Rozenberg, G. (Eds.), Vol. 5715, Springer, Berlin, **2009**.
9. *Unconventional Computing.* Adamatzky, A.; De Lacy Costello, B.; Bull, L.; Stepney, S.; Teuscher, C. (Eds.), Luniver Press, UK, **2007**.
10. De Silva, A. P.; Gunaratne, H. Q. N.; McCoy, C. P. *Nature* **1993**, *364*, 42-44.
11. De Silva, A. P.; Gunaratne, H. Q. N.; McCoy, C. P. *J. Am. Chem. Soc.* **1997**, *119*, 7891-7892.
12. De Silva, A. P.; Gunaratne, H. Q. N.; Maguire, G. E. M. *J. Chem. Soc., Chem. Commun.* **1994**, 1213-1214.
13. Baytekin, H. T.; Akkaya, E. U. *Org. Lett.* **2000**, *2*, 1725-1727.
14. Zong, G.; Xiana, L.; Lua, G. *Tetrahedron Lett.* **2007**, *48*, 3891-3894.
15. De Silva, A. P.; Dixon, I. M.; Gunaratne, H. Q. N.; Gunnlaugsson, T.; Maxwell, P. R. S.; Rice, T. E. *J. Am. Chem. Soc.* **1999**, *121*, 1393-1394.
16. Straight, S. D.; Liddell, P. A.; Terazono, Y.; Moore, T. A.; Moore, A. L.; Gust, D. *Adv. Funct. Mater.* **2007**, *17*, 777-785.
17. Turfan, B.; Akkaya, E. U. *Org. Lett.* **2002**, *4*, 2857-2859.
18. Wang, Z.; Zheng, G.; Lu, P. *Org. Lett.* **2005**, *7*, 3669-3672.
19. Gunnlaugsson, T.; Mac Dónaill, D. A.; Parker, D. *J. Am. Chem. Soc.* **2001**, *123*, 12866-12876.
20 Gunnlaugsson, T.; MacDónaill, D. A.; Parker, D. *Chem. Commun.* **2000**, 93-94.

21. De Sousa, M.; De Castro, B.; Abad, S.; Miranda, M. A.; Pischel, U. *Chem. Commun.* **2006**, 2051-2053.
22. Li, L.; Yu, M. X.; Li, F. Y.; Yi, T.; Huang, C. H. *Colloids Surf. A* **2007**, *304*, 49-53.
23. Credi, A.; Balzani, V.; Langford, S.J.; Stoddart, J.F. *J. Am. Chem. Soc.* **1997**, *119*, 2679-2681.
24. Li, Y.; Zheng, H.; Li, Y.; Wang, S.; Wu, Z.; Liu, P.; Gao, Z.; Liu, H.; Zhu, D. *J. Org. Chem.* **2007**, *72*, 2878-2885.
25. De Silva, A.P.; McClenaghan, N.D. *Chem. Eur. J.* **2002**, *8*, 4935-4945.
26. De Silva, A.P.; McClenaghan, N.D. *J. Am. Chem. Soc.* **2000**, *122*, 3965-3966.
27. Katz, E.; Privman, V. *Chem. Soc. Rev.* **2010**, *39*, 1835-1857.
28. Saghatelian, A.; Volcker, N. H.; Guckian, K. M.; Lin, V. S. Y.; Ghadiri, M. R. *J. Am. Chem. Soc.* **2003**, *125*, 346-347.
29. Ashkenasy, G.; Ghadiri, M. R. *J. Am. Chem. Soc.* **2004**, *126*, 11140-11141.
30. Sivan, S.; Lotan, N. *Biotechnol. Prog.* **1999**, *15*, 964-970.
31. Sivan, S.; Tuchman, S.; Lotan, N. *Biosystems* **2003**, *70*, 21-33.
32. Deonarine, A. S.; Clark, S. M.; Konermann, L. *Future Generation Computer Systems* **2003**, *19*, 87-97.
33. Ashkenazi, G.; Ripoll, D. R.; Lotan, N.; Scheraga, H. A. *Biosens. Bioelectron.* **1997**, *12*, 85-95.
34. Unger, R.; Moult, J. *Proteins* **2006**, *63*, 53-64.
35. Zhou, J.; Arugula, M. A.; Halámek, J.; Pita, M.; Katz, E. *J. Phys. Chem. B* **2009**, *113*, 16065-16070.
36. Zhou, J.; Tam, T. K.; Pita, M.; Ornatska, M.; Minko, S.; Katz, E. *ACS Appl. Mater. Interfaces* **2009**, *1*, 144-149.
37. Strack, G.; Ornatska, M.; Pita, M.; Katz, E. *J. Am. Chem. Soc.* **2008**, *130*, 4234-4235.
38. Pita, M.; Katz, E. *J. Am. Chem. Soc.* **2008**, *130*, 36-37.
39. Strack, G.; Chinnapareddy, S.; Volkov, D.; Halámek, J.; Pita, M.; Sokolov, I.; Katz, E. *J. Phys. Chem. B* **2009**, *113*, 12154-12159.
40. Tam, T. K.; Strack, G.; Pita, M.; Katz, E. *J. Amer. Chem. Soc.* **2009**, *131*, 11670-11671.
41. Li, T.; Wang, E. K.; Dong, S. J. *J. Am. Chem. Soc.* **2009**, *131*, 15082-15083.
42. Willner, I.; Shlyahovsky, B.; Zayats, M.; Willner, B. *Chem. Soc. Rev.* **2008**, *37*, 1153-1165.

43. Stojanovic, M. N.; Stefanovic, D.; LaBean, T.; Yan, H. in *Bioelectronics: From Theory to Applications*, Willner, I.; Katz, E. (Eds.), 427-455, Wiley-VCH, Weinheim, **2005**.
44. Ezziane, Z. *Nanotechnology* **2006**, *17*, R27-R39.
45. Win, M. N.; Smolke, C. D. *Science* **2008**, *322*, 456-460.
46. Rinaudo, K.; Bleris, L.; Maddamsetti, R.; Subramanian, S.; Weiss, R.; Benenson, Y. *Nature Biotechnol.* **2007**, *25*, 795-801.
47. Ogawa A.; Maeda, M. *Chem. Commun.* **2009**, 4666-4668.
48. Stojanovic, M. N.; Mitchell, T. E.; Stefanovic, D. *J. Am. Chem. Soc.* **2002**, *124*, 3555-3561.
49. Benenson, Y. *Curr. Opin. Biotechnol.* **2009**, *20*, 471-478.
50. Simpson, M. L.; Sayler, G. S.; Fleming, J. T.; Applegate, B. *Trends Biotechnol.* **2001**, *19*, 317-323.
51. Margulies, D.; Hamilton, A. D. *J. Am. Chem. Soc.* **2009**, *131*, 9142-9143.
52. May, E. E.; Dolan, P. L.; Crozier, P. S.; Brozik, S.; Manginell, M. *IEEE Sensors Journal* **2008**, *8*, 1011-1019.
53. Wang, J.; Katz, E. *Anal. Bioanal. Chem.* **2010**, in press (DOI: 10.1007/s00216-010-3746-0).
54. Strack, G.; Bocharova, V.; Arugula, M. A.; Pita, M.; Halámek, J.; Katz, E. *J. Phys. Chem. Lett.* **2010**, *1*, 839-843.
55. Tokarev, I.; Gopishetty, V.; Zhou, J.; Pita, M.; Motornov, M.; Katz, E.; Minko, S. *ACS Appl. Mater. Interfaces* **2009**, *1*, 532-536.
56. Simmel, F. C. *Nanomedicine* **2007**, *2*, 817-830.
57. von Maltzahn, G.; Harris, T. J.; Park, J.-H.; Min, D.-H.; Schmidt, A. J.; Sailor, M. J.; Bhatia, S. N. *J. Am. Chem. Soc.* **2007**, *129*, 6064-6065.
58. Manesh, K. M.; Halámek, J.; Pita, M.; Zhou, J.; Tam, T. K.; Santhosh, P.; Chuang, M.-C.; Windmiller, J. R.; Abidin, D.; Katz, E.; Wang, J. *Biosens. Bioelectron.* **2009**, *24*, 3569-3574.
59. Pita, M.; Zhou, J.; Manesh, K. M.; Halámek, J.; Katz, E.; Wang, J. *Sens. Actuat. B* **2009**, *139*, 631-636.
60. Strack, G.; Pita, M.; Ornatska, M.; Katz, E. *ChemBioChem* **2008**, *9*, 1260-1266.
61. Baron, R.; Lioubashevski, O.; Katz, E.; Niazov, T.; Willner, I. *J. Phys. Chem. A* **2006**, *110*, 8548-8553.

62. Niazov, T.; Baron, R.; Katz, E.; Lioubashevski, O.; Willner, I. *Proc. Natl. Acad. USA.* **2006**, *103*, 17160-17163.

63. Qu, D.-H.; Wang, Q.-C.; Tian, H. *Angew. Chem. Int. Ed.* **2005**, *44*, 5296-5299.

64. Andréasson, J.; Straight, S. D.; Kodis, G.; Park, C.-D.; Hambourger, M.; Gervaldo, M.; Albinsson, B.; Moore, T. A.; Moore, A. L.; Gust, D. *J. Am. Chem. Soc.* **2006**, *128*, 16259-16265.

65. Andréasson, J.; Kodis, G.; Terazono, Y.; Liddell, P. A.; Bandyopadhyay, S.; Mitchell, R. H.; Moore, T. A.; Moore, A. L.; Gust, D. *J. Am. Chem. Soc.* **2004**, *126*, 15926-15927.

66. Lopez, M. V.; Vazquez, M. E.; Gomez-Reino, C.; Pedrido, R.; Bermejo, M. R. *New J. Chem.* **2008**, *32*, 1473-1477.

67. Margulies, D.; Melman, G.; Shanzer, A. *J. Am. Chem. Soc.* **2006**, *128*, 4865-4871.

68. Kuznetz, O.; Salman, H.; Shakkour, N.; Eichen, Y.; Speiser, S. *Chem. Phys. Lett.* **2008**, *451*, 63-67.

69. Ho, M. S. H.; Lin, J.; Chiu, C. C. *International Journal of Innovative Computing Information and Control* **2009**, *5*, 4573-4582.

70. Baron, R.; Lioubashevski, O.; Katz, E.; Niazov, T.; Willner, I. *Angew. Chem. Int. Ed.* **2006**, *45*, 1572-1576.

71. Donald, J.; Jha, N. K. *ACM Journal on Emerging Technologies in Computing Systems* **2008**, *4*, art. #2.

72. Shende, V. V.; Markov, I. L. *Quantum Information & Computation* **2009**, *9*, 461-486.

73. Bennett, C. H. *IBM J. Res. Developm.* **1973**, *17*, 525-532.

74. Ekert, A.; Jozsa, R. *Rev. Mod. Phys.* **1996**, *68*, 733-753.

75. Steane, A. M. *Rept. Prog. Phys.* **1998**, *61*, 117-173.

76. Fredkin, E.; Toffoli, T. *Int. J. Theor. Phys.* **1982**, *21*, 219-253.

77. *Methods of Enzymatic Analysis*, Vol. 4, 2<sup>nd</sup> ed., Bergmeyer, H. U. (Ed.), Academic Press, New York, **1974**, pp. 2066–2072.

78. Privman, V.; Strack, G.; Solenov, D.; Pita, M.; Katz, E. *J. Phys. Chem. B* **2008**, *112*, 11777-11784.

79. Melnikov, D.; Strack, G.; Pita, M.; Privman, V.; Katz, E. *J. Phys. Chem. B* **2009**, *113,* 10472-10479.

80. Privman, V.; Pedrosa, V.; Melnikov, D.; Pita, M.; Simonian, A.; Katz, E. *Biosens. Bioelectron.* **2009**, *25*, 695-701.
81. Fedichkin, L.; Katz, E.; Privman, V. *J. Comput. Theor. Nanoscience* **2008**, *5*, 36-43.
82. Soldin, S. J.; Balinsky, D. *Biochemistry* **1968**, *7*, 1077-1082.
83. Shreve, D. S.; Levy, H. R. *J. Biol. Chem.* **1980**, *255*, 2670-2677.
84. Özer, N.; Aksoy, Y.; Öğüs, I. H. *Int. J. Biochem. Cell Biol.* **2001**, *33*, 221-226.
85. Scopes, R. K. *Biochem. J.* **1997**, *326*, 731-735.
86. Privman, V.; Arugula, M. A.; Halámek, J.; Pita, M.; Katz, E. *J. Phys. Chem. B* **2009**, *113,* 5301-5310.
87. Nazari, K.; Mahmoudi, A.; Khosraneh, M.; Haghighian, Z.; Moosavi-Movahedi, A. A. *J. Molec. Catal. B* **2009**, *56*, 61-69.
88. Khosraneh, M.; Mahmoudi, A.; Rahimi, H.; Nazari, K.; Moosavi-Movahedi, A.A. *Journal of Enzyme Inhibition and Medicinal Chemistry* **2007**, *22*, 677-684.

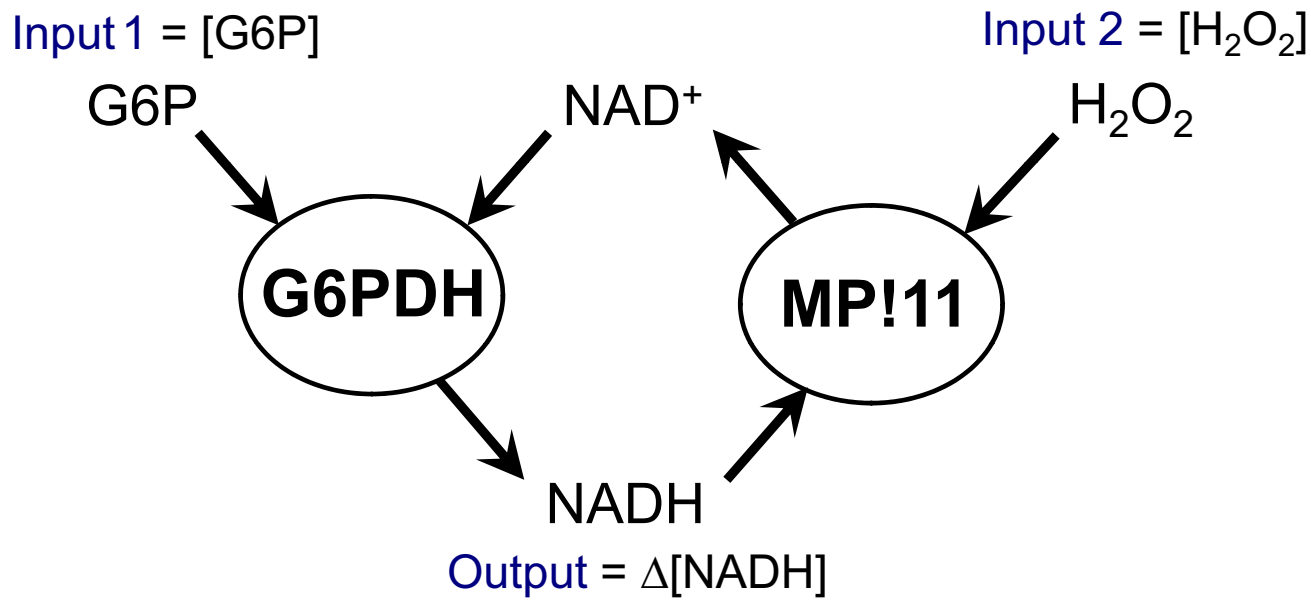

**Scheme 1.** Schematic of the biocatalytic processes utilized in the **XOR** gate functioning. (The abbreviations for the chemicals are defined in the text.) *Top Inset:* The truth table of the **XOR** gate.

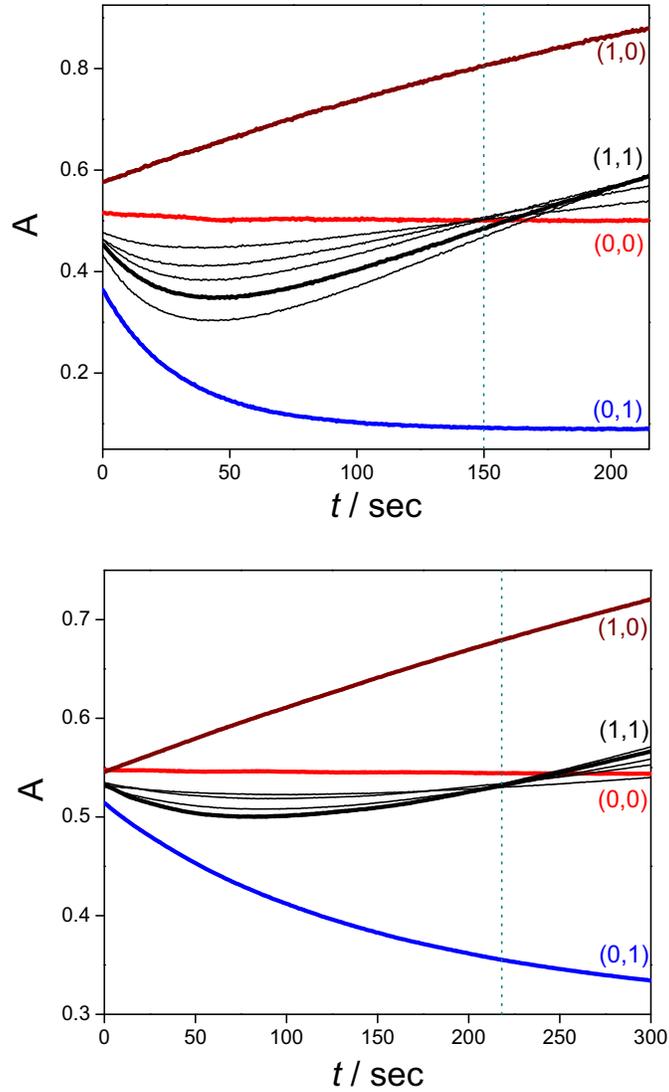

**Figure 1.** The four heavy-symbol lines show the time-dependent data obtained with inputs at the selected referenced logic-**0** or **1** values (four possible combinations). The top panel corresponds to the original biocatalyst concentrations, whereas the bottom panel gives data for the optimized-gate case (with reduced noise amplification), as detailed in Section 4. Out of the total 36 data sets obtained by varying the inputs independently in ratios 0, 1/4, 1/2, 3/4 1, 5/4 of the logic-**1** values, we show, in addition to the 4 combinations (0,0), (0,1), (1,0), (1,1), also the curves obtained for inputs at (1/4,1/4), (1/2,1/2), (3/4,3/4), (5/4,5/4) of the logic-**1** inputs. These data are useful in explaining the selections of the sampling (gate) times (see text), marked by dotted vertical lines. The four additional time-dependent data sets are plotted with less-heavy, black symbols, and their fixed-time values are monotonically decreasing at times close to 50 sec in the top panel; monotonically increasing at times close to 300 sec in the bottom panel.

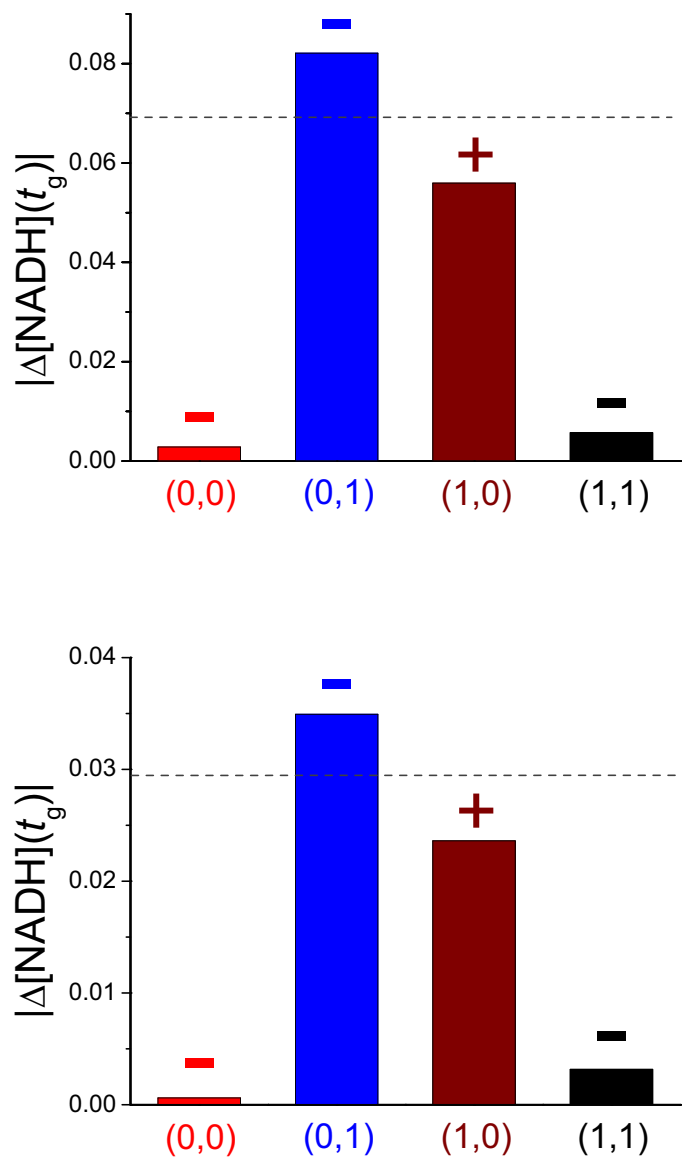

**Figure 2.** Bar charts illustrating the degree of accuracy to which the **XOR** function is realized in the present system. The signals were obtained from the absorbance (Figure 1) data at the selected $t_g$ values, as the absolute values of the concentration differences — with respect to the $t = 0$ logic input (**0,0**) reference — for NADH. The signs of the concentration differences before taking the absolute value are also shown. The values are color-coded to the data in Figure 1, and the top panel corresponds to the original biocatalyst concentrations (as in Figure 1), whereas the bottom panel corresponds to the modified experiment. The dashed horizontal lines show the selected logic-**1** values, as explained in the text: 0.069 mM and 0.029 mM, for the original and modified experiments, respectively.

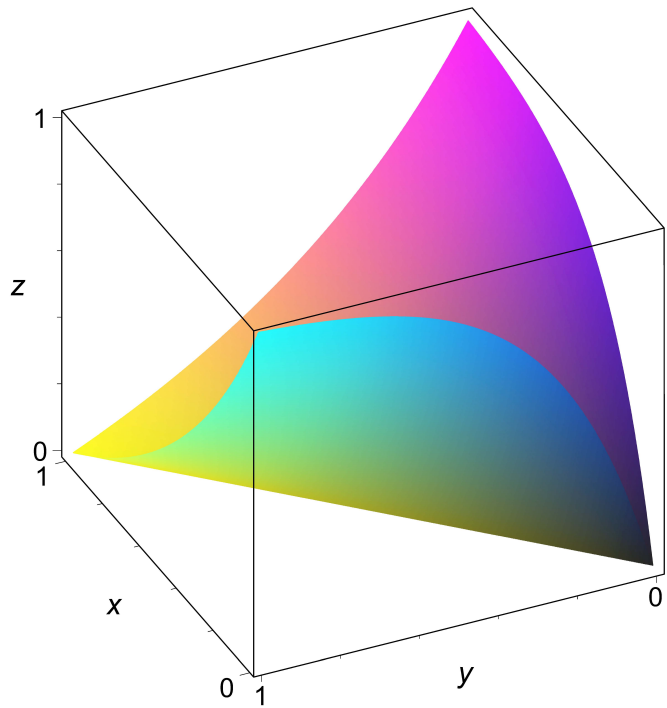

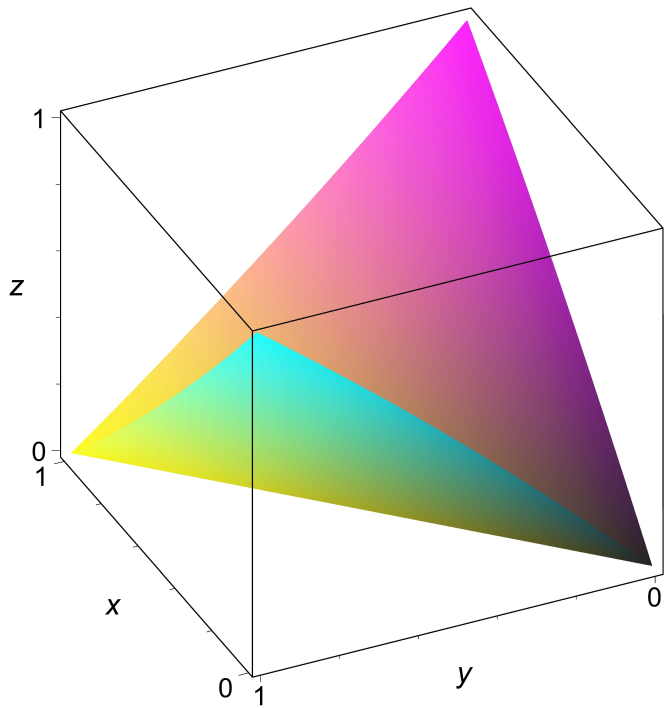

**Figure 3.** The shape of the model response function for our **XOR** gate, with *a* = 4 (*top panel*) and ¼ (*bottom panel*). Note that the origin of the three-dimensional coordinate system is in the lower-right corner of the unit cube in this view.

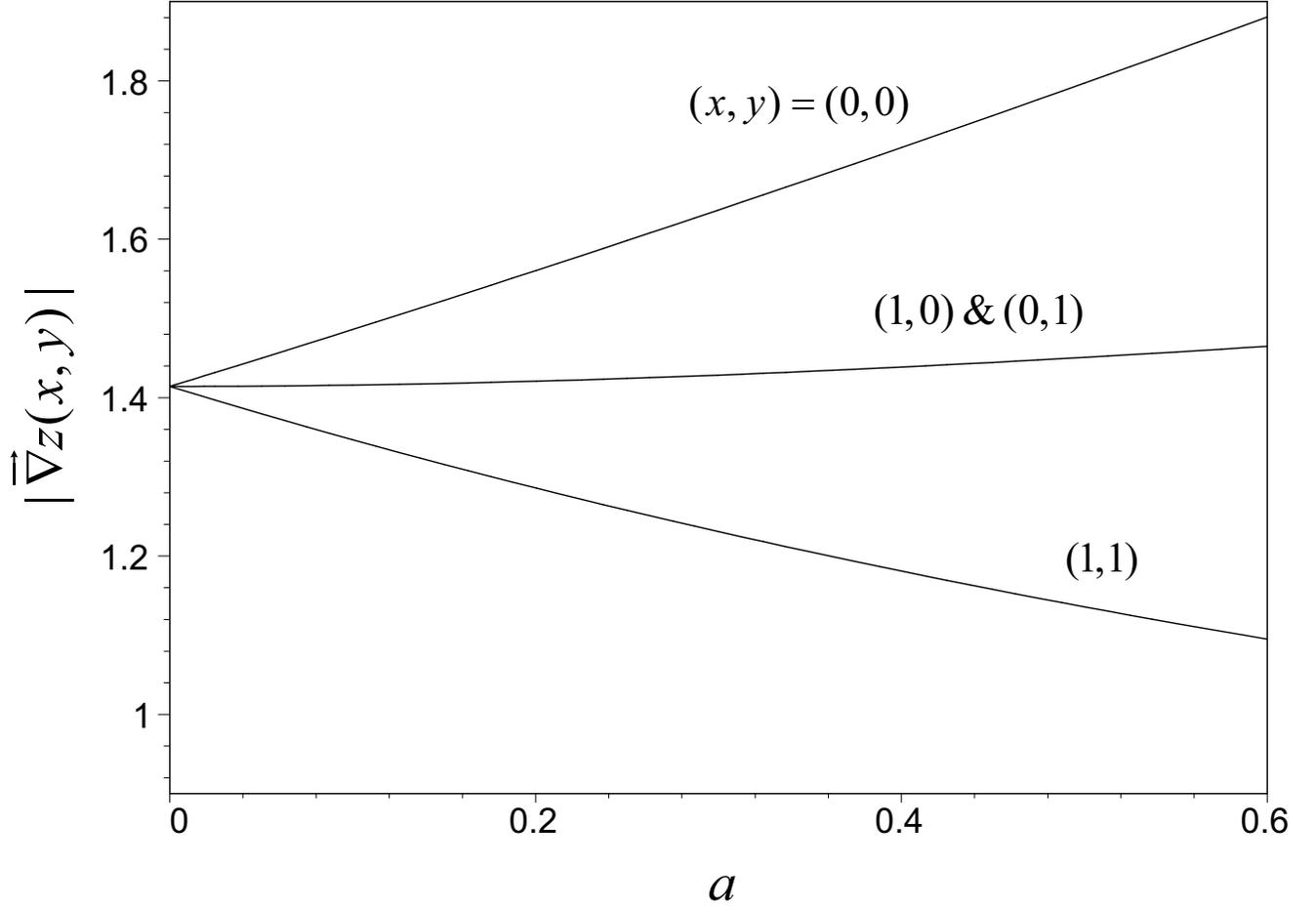

**Figure 4.** Variation of the slopes of the gate-response function in Eq. (6) with the parameter *a*. The curves give the magnitudes of the gradients at the four logic points, as marked in the plot. For *x* = *y*, the gradients were calculated before taking the absolute value in Eq. (6).

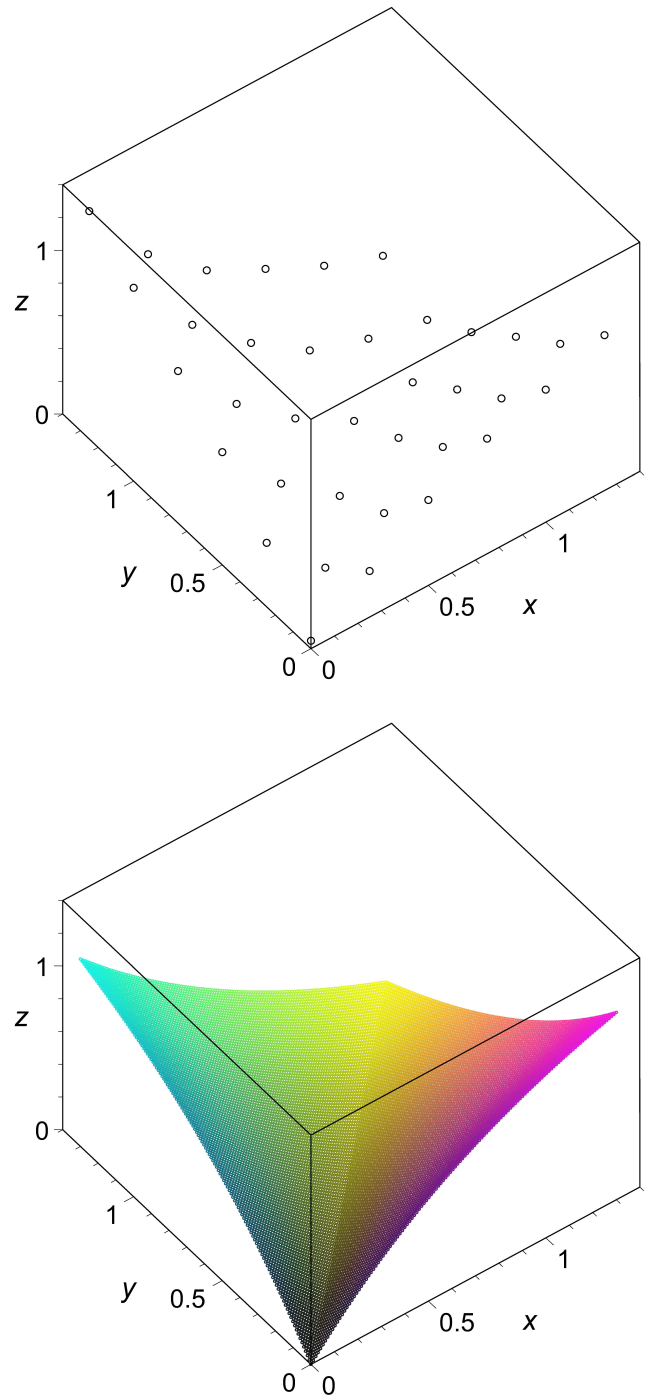

**Figure 5.** *Top panel:* Experimental data scaled by the "logic-**1**" output value, for the original data set. *Bottom panel:* Model fit of the data, yielding $a = 1.06$. Note that both the data and the fitting function, Eq. (6), here extend beyond the logic-**1** values.

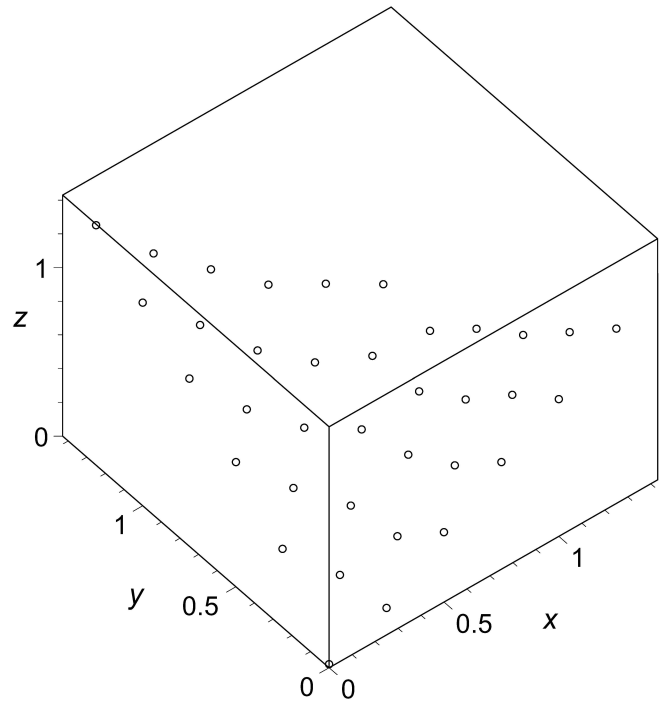

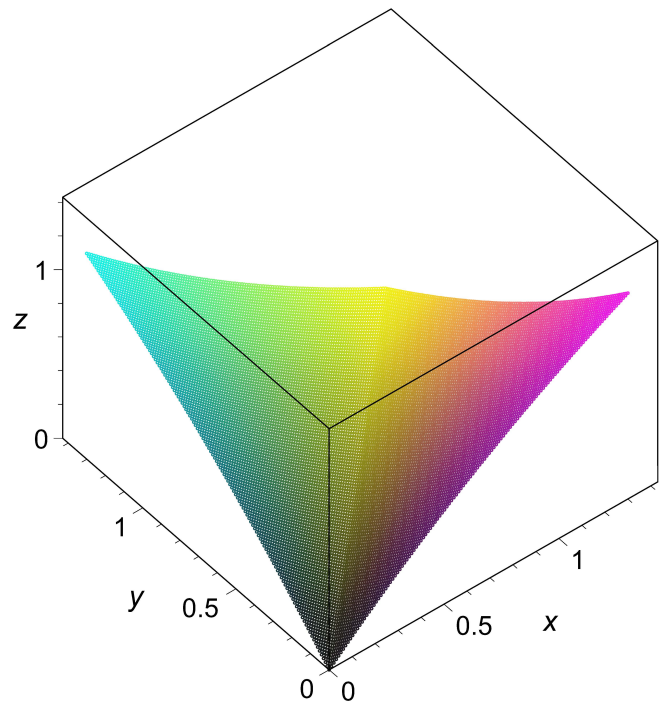

**Figure 6.** Same as Figure 5, but for the "optimized experiment" data set, with the fitted value *a* = 0.459.